\documentclass[
  journal=pasa,
  manuscript=research-paper,
  year=202X,
  volume=XX,
]{cup-journal}

\usepackage{amsmath}
\usepackage{amssymb}
\usepackage{hyperref}
\usepackage[nopatch]{microtype}
\usepackage{booktabs}
\usepackage{cuted}

\title{Improved Modeling for Moving Sources of Radio Frequency Interference and Impact on Flagging Strategies}

\author{Jade M. Ducharme}
\affiliation{Department of Physics, Brown University, Providence, RI 02912, USA}
\email[Jade M. Ducharme]{jade\_ducharme@brown.edu}

\author{Jonathan C. Pober}
\affiliation{Department of Physics, Brown University, Providence, RI 02912, USA}

\author{Michael J. Wilensky}
\affiliation{Department of Physics and Trottier Space Institute, McGill University, 3600 University Street, Montreal, QC H3A 2T8, Canada}
\alsoaffiliation{CITA National Fellow}


\begin{document}

\begin{abstract}
Radio frequency interference (RFI) is a major challenge for low-frequency radio experiments targeting the redshifted 21-cm signal from the Epoch of Reionization. Many important RFI sources, including aircraft and satellites, move across the sky during an observation. The averaging of visibilities for such moving sources over a finite correlator integration produces a distinct sinc-like pattern in the interferometric $uv$-plane, causing the RFI to appear bright on some baselines and strongly suppressed on others. Here, we develop an analytical model for this effect and validate its qualitative behavior against real MWA observations and numerical simulations. Controlled RFI injections into clean MWA data are then used to compare several flagging treatments in the context of EoR power spectrum recovery. These treatments explore the trade-off between per-baseline flagging, which can miss weak but potentially non-negligible contamination on baselines where the integration effect leads to strong suppression, and baseline-aggregated flagging, which removes more faint residual contamination but reduces the total number of usable $uv$-modes.  We find that the preferred strategy depends on RFI brightness and occupancy, but that flagging all frequencies at contaminated time-steps most consistently minimizes excess power for bright or frequent events. However, none of the tested flagging treatments fully recovers the uncontaminated reference spectrum, motivating future work on direct modeling and subtraction of moving RFI sources.
\end{abstract}

\section{Introduction}

Radio frequency interference (RFI) is an increasingly pressing challenge for radio astronomy operations. Epoch of Reionization (EoR) experiments, which demand above all precise instrumentation and pristine observing conditions, are particularly impacted by the growing number of objects contributing to signal contamination, including airplanes, satellites, and satellite constellations \citep{Wilensky_2020, Wilensky_2023, DiVruno_2023, Grigg_2023, starlink3, Grigg_2025}. 

A common feature of many important RFI sources is that they move across the sky during an observation. In this work, we show that this motion produces a predictable and analytically tractable signature in the interferometric $uv$-plane. Unlike a stationary source, which yields approximately uniform visibility amplitudes across baselines, a moving source---which travels over the course of a single correlator integration---averages to produce a sinc-like interference pattern in $uv$-space, causing the RFI to appear brighter on some baselines and fainter on others.

This baseline-dependent effect has direct consequences for RFI mitigation, typically handled through offline flagging. This method examines individual visibilities and determines which samples should be flagged as contaminated---thereby removing them from further analysis. Offline analyses generally work in the time-frequency domain, where the visibility data from each interferometric baseline (pair of antennas) and polarization are represented as two-dimensional time vs. frequency ``waterfall'' plots. In this space, RFI contamination, which typically spans a finite bandwidth and duration, appears as a bright rectangle or streak.

Although bright RFI can often be identified by eye in a waterfall plot, designing reliable automated flagging algorithms remains an active area of research and a central focus for many radio interferometers targeting the EoR or post-reionization eras, such as the Murchison-Widefield Array (MWA\endnote{\url{https://www.mwatelescope.org/}}; \citealt{mwa1, mwa2}), the Hydrogen Epoch of Reionization Array (HERA\endnote{\url{https://reionization.org/}}; \citealt{DeBoer_2017}), the LOw-Frequency ARray (LOFAR\endnote{\url{https://www.astron.nl/telescopes/lofar/}}; \citealt{lofar}), and the Canadian Hydrogen Intensity Mapping Experiment (CHIME\endnote{\url{https://chime-experiment.ca/}}; \citealt{chime}). While our analysis uses the MWA as a specific example, we discuss the applicability of our main conclusions to other experiments searching for the faint 21-cm power spectrum in Section \ref{sec:discussion}.

For the MWA, commonly used flagging tools include \texttt{AOFlagger} \citep{AOFlagger}, \texttt{SSINS} \citep{Wilensky_2019}, and the redundant-calibration-based $\chi^2$ flagger \citep{Kunicki_Pober_2024}. These methods differ in how they use information across baselines. Per-baseline approaches, such as the MWA's default implementation of \texttt{AOFlagger}, examine each baseline independently and produce a separate flag mask for each baseline. Baseline-aggregated approaches, such as  \texttt{SSINS} and $\chi^2$, combine information across baselines to construct a single time-frequency statistic, from which a global flag mask is derived and applied to all baselines.

The $uv$-space structure of moving RFI raises an important question for flagging strategies. Because, as we shall show, the moving-source signal can be bright on some baselines and nearly invisible on others, per-baseline flaggers may detect and flag contamination only on the baselines where it is strongest. Baseline-aggregated methods, on the other hand, take advantage of the most highly contaminated baselines when constructing their global flag mask, allowing RFI on even the most suppressed baselines to be flagged. However, more aggressive flagging is not automatically better for EoR science: removing data reduces $uv$-coverage, while frequency-dependent flag masks introduce spectral structure \citep{wilensky_2022}. Both effects can introduce excess power in an EoR power spectrum analysis. Alternatively, some experiments, such as LOFAR, flag RFI at a higher spectral resolution before averaging down to the required power spectrum resolution (ignoring flagged samples in the averaging), but this method can also lead to excess power \citep{offringa_2019}. The relevant question is therefore not simply which method detects the most RFI, but which method produces the least biased power spectrum.

In this work, we address this question in two steps. First, we develop an analytical model for the $uv$-plane response of a moving RFI source and validate it against both real MWA data and numerical simulations. Second, we use controlled injection experiments, combining real MWA observations with simulated moving RFI, to compare several flagging treatments in the context of EoR power spectrum recovery. These treatments include leaving the RFI unflagged, flagging the contaminated time-frequency region on all baselines, flagging all frequencies at contaminated time-steps, and applying \texttt{AOFlagger} on a per-baseline basis.

Beyond comparing existing flagging strategies, the analytical model developed here also points toward a longer-term alternative: direct RFI modeling and subtraction. Previous work has focused primarily on localizing RFI sources in image space \citep{prabu2023, Ducharme_Pober_2025}. By contrast, our framework describes the expected structure of moving-source contamination directly in visibility space. This is important because, as we show, even the best-performing flagging treatment tested here remains measurably biased relative to an uncontaminated data set. Flagging can reduce RFI contamination, but it cannot recover information lost through data excision.

The remainder of this paper is organized as follows. In Section \ref{sec:stationary_vs_moving}, we present two MWA observations containing clear instances of RFI and show that moving emitters produce distinct signatures in the $uv$-plane. In Section \ref{sec:modeling}, we develop a fully analytical framework that explains these patterns from first principles and compare it to numerical simulations. Section \ref{sec:flagging} examines the implications of this behaviour for RFI flagging strategies, with Sections \ref{sec:experimental_design} through \ref{sec:benchmark} focusing specifically on their impact on EoR power spectrum recovery. In Section \ref{sec:discussion}, we discuss the results and implications of our power spectrum experiments. Finally, we summarize our key findings in Section \ref{sec:conclusion}.

\section{Stationary vs. Moving RFI Sources in $uv$-Space} \label{sec:stationary_vs_moving}

Interferometric visibilities are cross-correlations of the voltages measured by pairs of antennas, recorded as functions of baseline, time, frequency, and polarization. Each baseline is described by a vector separation between antennas, expressed in the $(u,v,w)$ coordinate system. In this coordinate system, wherein distances are measured in units of wavelengths, the $uv$-plane is normal to the line-of-sight $w$-axis. When considering a small enough field-of-view, the flat sky approximation applies, and the $w$-axis can be largely ignored. 

Visibilities are complex-valued: for signals from a given direction, the finite separation between antennas introduces a geometric path delay, which is encoded in the visibility phase. A stationary source produces an approximately fixed geometric delay for a given baseline, time, and frequency combination. Geometrically, a fixed source and two antennas define a configuration that barely changes over a typical correlator integration time (order seconds), so the corresponding visibility phase remains coherent over that interval.

A moving source, on the other hand, produces a range of delays within a typical correlation interval. As its apparent position changes, the source-antenna geometry evolves, causing the visibility phase to vary across the integration window. This leads to a partial decoherence of the signal and the emergence of a distinct pattern in the $uv$-plane.

To illustrate these motion-dependent effects, we select two MWA Phase II observations chosen to be representative examples. Both of them are found to contain RFI via visual inspection of single-baseline waterfall plots, examples of which are shown in the top row of Figure \ref{fig:compare_stationary_moving}. The first one (OBSID 1160763448; taken on 2016-10-17 at 18:17 UTC) is confirmed through imaging to contain a moving source of RFI, as shown in the see middle-left panel of Figure \ref{fig:compare_stationary_moving}). By contrast, imaging of the second observation (OBSID 1161716960; taken on 2016-10-28 at 19:09 UTC), shown in the middle-right panel of Figure \ref{fig:compare_stationary_moving}, reveals no dominant source of RFI. It is worth noting that OBSID 1161716960 is thought to be contaminated by RFI from a beyond-the-horizon digital television transmitter that reached the array through tropospheric ducting. Similar events will be discussed in more detail in forthcoming work \citep{EAVILS}.

\subsection{Data Pre-processing}\label{sec:preprocessing}

Prior to any analysis, OBSID 1160763448 was calibrated with the \texttt{Hyperdrive} software \citep{hyperdrive} with default settings. To prevent RFI contamination from biasing the calibration solutions, flags were applied; specifically, we used \texttt{SSINS}, which is the current default flagging strategy for the MWA EoR project\footnote{This work does not explore second-order effects related to calibration bias introduced by flagging strategies. We limit our scope to the effects of flagging and of residual unflagged RFI on a power spectrum analysis.}.

For OBSID 1161716960, since RFI contaminates a specific frequency band throughout the entire observation, it was impossible to obtain reliable calibration solutions for those frequencies. As a practical alternative, we transferred calibration solutions from a nearby observation. Because this particular RFI event persisted throughout the entire night, no suitable clean observation was available from the same night. We therefore used OBSID 1161863064, taken roughly 40 hours later, and calibrated it using the default settings for \texttt{Hyperdrive}. MWA antenna gains are generally stable over such timescales \citep{star2024}, providing sufficiently reliable transferred calibration solutions.

Tile quality assurance was then assessed for both observations following the approach of \cite{Nunhokee2024}, which involves examining time-averaged autocorrelations, normalized to a reference antenna, to identify antennas with anomalous bandpass gains prior to calibration. Antennas with modified z-scores exceeding 3.5 were identified as outliers and excluded from the calibration and from all subsequent analysis. 

Finally, a sky model was subtracted from both observations, again using \texttt{Hyperdrive}, in order to fully isolate the RFI signals. This is done using the default \texttt{Hyperdrive} settings.

In both observations, we assume the RFI is confined to the frequency channels corresponding to the Australian DTV-7 allocation, spanning the range between 181 and 188 MHz \citep{dtv_allocations}. This assumption is supported by visual inspection of contaminated waterfall plots. Accordingly, after calibration and imaging, we restrict the remainder of our analysis to this band by averaging all visibilities within 181-188 MHz and discarding data outside this frequency range.

\subsection{$uv$-Plane Signatures}

\begin{figure*}[hbt!]
    \centering
    \includegraphics[width=0.9\linewidth]{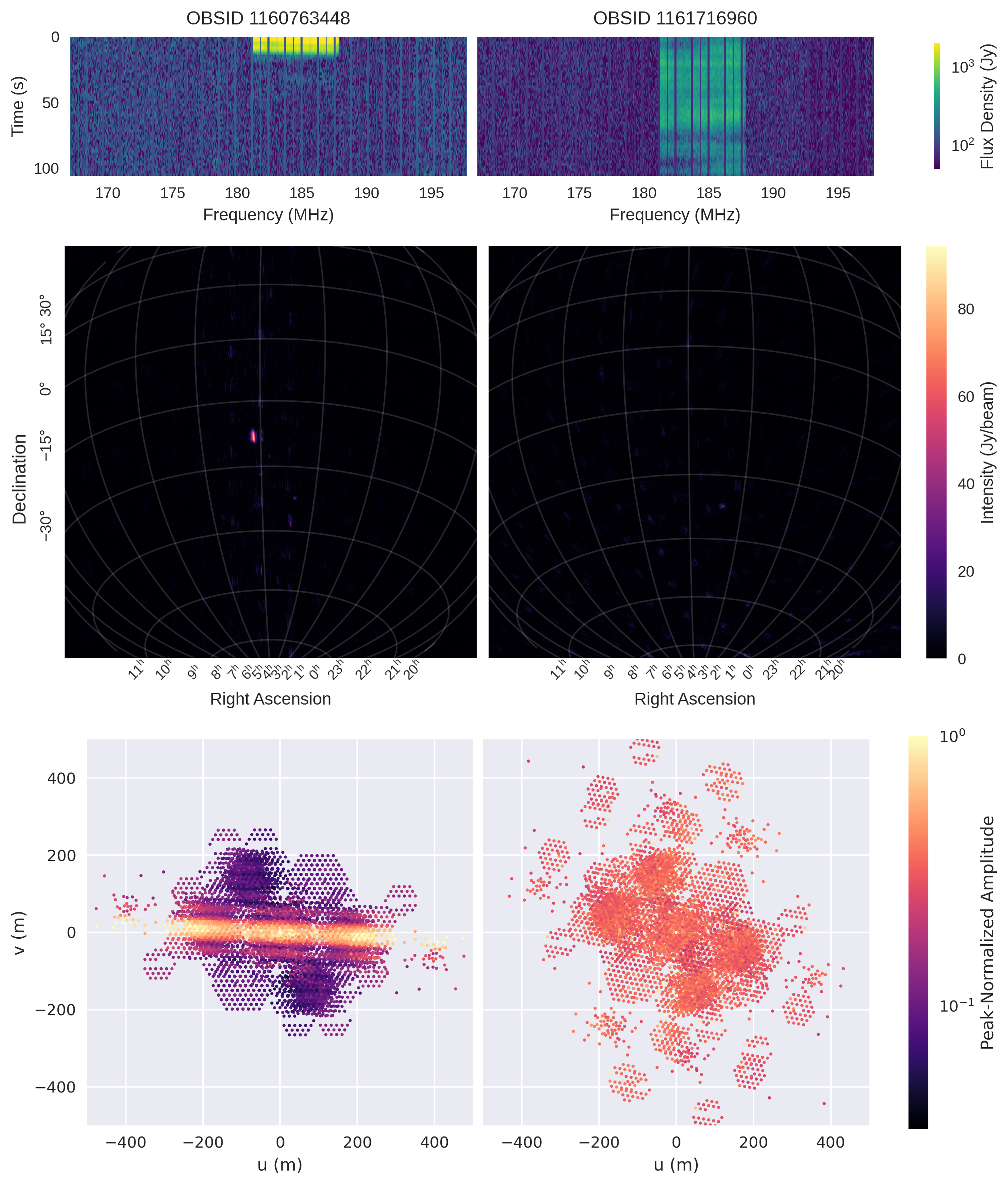}
    \caption{Top row: single-baseline waterfall plots for OBSIDs 1160763448 and 1161716960 (after calibration and sky-subtraction). Middle row: the most heavily contaminated time integrations (i.e., those with the highest baseline-averaged visibility amplitude) for each OBSID, imaged across the DTV-7 frequency band using \texttt{WSClean}. Specifically, the first time integration is imaged for OBSID 1160763448, while for 1161716960, we image 62 seconds into the observation. The first image clearly resolves a source of RFI, whereas the second does not. Note that the source around right ascension 3h22 and declination -37$^\circ$ in both images is a residual of Fornax A and not associated with RFI. Bottom row: for each OBSID, data are restricted to the most contaminated time integration and to the DTV-7 band; the resulting frequency-averaged visibility amplitudes are then plotted for each baseline at its corresponding $uv$-coordinate. The moving source on the left produces a striking sinc-like pattern. By contrast, OBSID 1161716960 shows an approximately constant response across baselines. OBSID 1161716960 includes a larger number of baselines because the tile quality assurance test (see Section \ref{sec:preprocessing}) did not flag the same set of tiles in both observations.}
    \label{fig:compare_stationary_moving}
\end{figure*}

To select a representative example of strong RFI contamination, we identify the time integration with the highest baseline-averaged visibility amplitude in each observation. For OBSIDs 1160763448 and 1161716960 respectively, this integration is located 0 and 62 seconds into the observation. The visibility amplitudes at these selected integrations are then plotted for each baseline at its corresponding $uv$-coordinate (autocorrelations excluded). This is shown in the bottom row of Figure \ref{fig:compare_stationary_moving}.

\begin{figure*}[h!]
    \centering
    \includegraphics[width=0.9\linewidth]{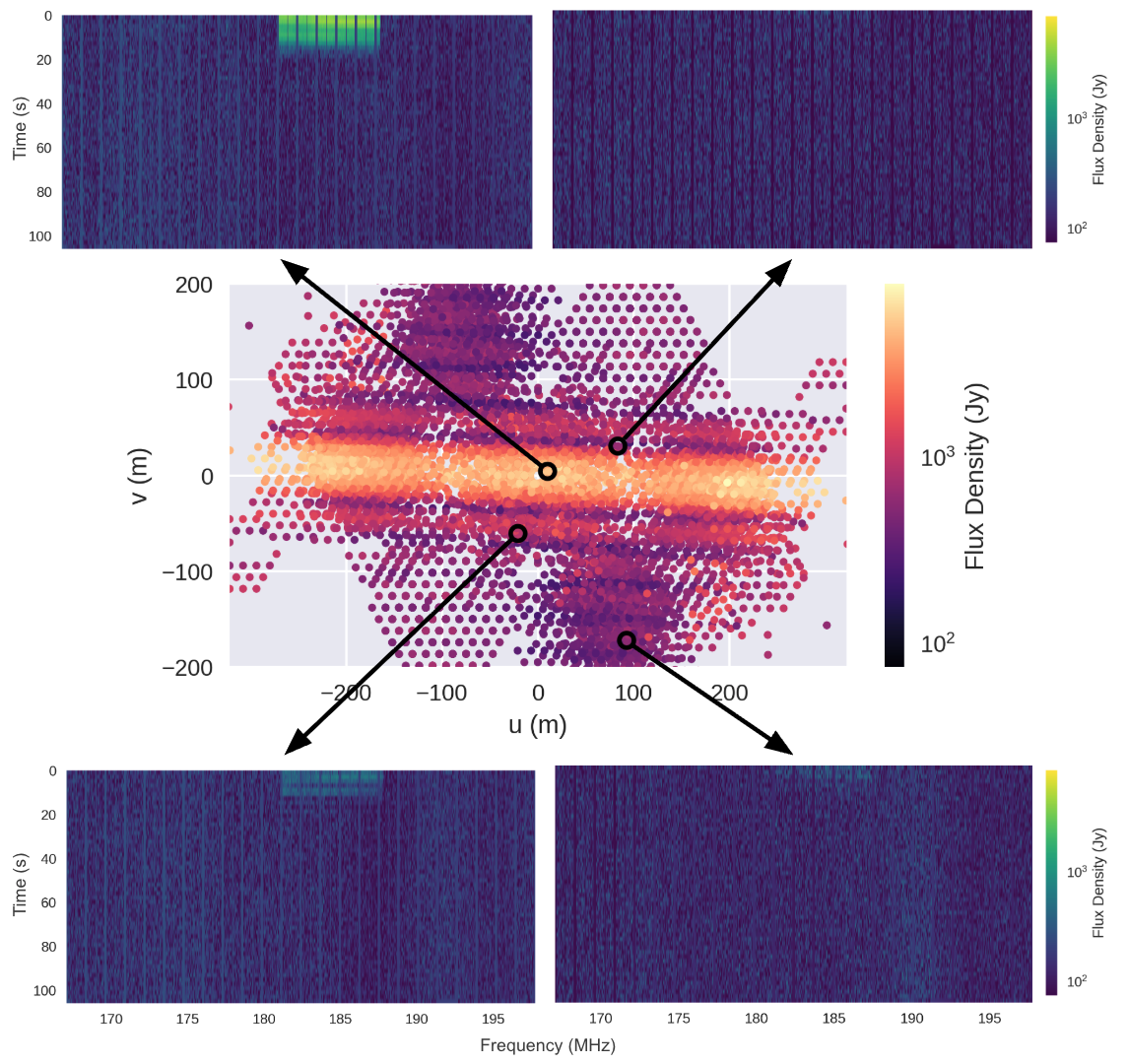}
    \caption{OBSID 1160763448: Waterfall plots for four select baselines are shown in the four corners (after calibration and sky-subtraction). The central panel shows each baseline's $uv$-coordinates and its average visibility amplitude across the DTV-7 band for the first time integration. Baselines within the bright central streak of the sinc-like pattern present strong RFI features in their waterfall plots (e.g.; top-left panel), whereas baselines in the suppressed nulls of the pattern show little to no detectable RFI (e.g.; top-right panel).}
    \label{fig:wf_vs_uv_moving}
\end{figure*}

We note the differences in the distribution of visibility amplitudes in the $uv$-plane for both cases. For the unresolved source of RFI, all baselines record a comparable signal amplitude; for the moving source of RFI, however, a clear sinc-like pattern emerges. In practice, this leads to RFI appearing exceptionally bright on baselines that fall within the peaks of the pattern, while those in the troughs see the signal largely suppressed. Figure \ref{fig:wf_vs_uv_moving} illustrates this effect with several single-baseline waterfall plots drawn from different baselines for OBSID 1160763448. 

The width and orientation of this pattern depend on the specific trajectory and speed of the moving source. \ref{appendix:uv_plane} presents additional MWA observations featuring a range of moving RFI events, illustrating both the recurrence of this behaviour and the variation in the resulting patterns.

\section{Modeling Interferometric Visibilities} \label{sec:modeling}

As discussed in Section \ref{sec:stationary_vs_moving}, an interferometer does not directly produce an image of the sky but instead measures visibilities, which are defined in the $uv$ coordinate system. In the flat-sky limit, performing an inverse Fourier transform over visibilities carries these measurements into their Fourier-dual sky-coordinate system, enabling the reconstruction of a sky image.

It follows that this process can also be inverted. Starting with a known sky signal, one can perform a forward Fourier transform in order to generate visibilities. In the sub-sections that follow, we derive a fully analytical model that reproduces the $uv$-space structure of moving RFI sources and compare its predictions with those from simulation tools.

\subsection{Analytical Model}\label{sec:analytical_model}

While a stationary source can be represented by a delta function in image space, a moving RFI emitter is better approximated as a thin streak. Indeed, over the typical two-second integration used in the observations considered in Section \ref{sec:stationary_vs_moving}, motion from an RFI source such as an airplane (around one degree per second; \citealt{Ducharme_Pober_2025}), would cause the signal to ``smear'' into a narrow streak across the sky.

Since RFI sources typically travel at roughly constant altitudes, we can describe them within a two-dimensional unitless Cartesian $\ell m$-plane on the sky, where $\ell$ and $m$ are mutually orthogonal and both perpendicular to the line-of-sight $n$-axis. These are called the \textit{direction cosines}, used to locate sources on a unit sphere centred on the observer. The streak can then be characterized via Equation \ref{eq:streak_eqn}.

\begin{equation}\label{eq:streak_eqn}
        f(\ell,m)=
        \begin{cases}
        1, & \left| s \right| \le \frac{L}{2},\quad \left| t \right| \le \frac{W}{2},\\
        0, & \text{otherwise,}
        \end{cases}
    \end{equation}
    with
    \begin{equation*}
    \begin{split}
        s &= (\ell-\ell_0)\cos\theta + (m-m_0)\sin\theta,\\
        t &= -(\ell-\ell_0)\sin\theta + (m-m_0)\cos\theta,
    \end{split}
    \end{equation*}
where $(\ell_0, m_0)$ are the coordinates of the streak's starting position, $L$ is its length, $\theta$ is the streak's angle with respect to the $\ell$-axis, and $W$ is its width.

Imaging software such as \texttt{WSClean} \citep{offringa-wsclean-2014, offringa-wsclean-2017} may prefer to locate sources using right ascension and declination. Since this work makes use of \texttt{WSClean} for the purpose of source finding in Section \ref{sec:comparison_with_simulation}, it is necessary to define the conversion from right ascension and declination coordinates to the direction cosine system, performed via the transformation presented in Equation \ref{lm_ra_dec_conversion} \citep{Thompson2001}.

\begin{equation} \label{lm_ra_dec_conversion}
    \begin{aligned}
        \ell & =\cos \delta \sin (\alpha-\alpha_0), \\
        m & =\sin \delta \cos \delta_0-\cos \delta \sin \delta_0 \cos (\alpha-\alpha_0),
    \end{aligned}
\end{equation}
where $\alpha$ and $\delta$ correspond respectively to the right ascension and declination coordinates of the source, and $\alpha_0$ and $\delta_0$ to the right ascension and declination of the phase centre of the observation.

Having thus obtained an analytical representation of a moving source of RFI in image space, we turn our attention to its representation in $uv$-space. In order to convert between the two, we apply a two-dimensional Fourier transform, as defined in Equation \ref{eq:fourier_2d} \citep{Thompson2001}.

\begin{equation} \label{eq:fourier_2d}
    F(u, v)=\int_{-\infty}^{\infty} \int_{-\infty}^{\infty} f(\ell, m) e^{-i 2 \pi(u \ell +v m)} d \ell d m,
\end{equation}

where $f(\ell, m)$ is the sky signal and $F(u, v)$ generates the visibilities in the $uv$-plane. For the two-dimensional streak defined in Equation \ref{eq:streak_eqn}, the Fourier transform provides the analytical solution presented in Equation \ref{eq:analytical_ft_solution}.
 \begin{equation}\label{eq:analytical_ft_solution}
        F(u,v) = L \, W
        \operatorname{sinc}(L  u_\parallel)\,
        \operatorname{sinc}(W u_\perp) \, e^{-i2\pi(u\ell_0+vm_0)},
    \end{equation}
    with
    \begin{equation*}
        \begin{split}
            u_\parallel &= u\cos\theta + v\sin\theta,\\
            u_\perp &= -u\sin\theta + v\cos\theta ,
        \end{split}
    \end{equation*}
    and
    \begin{equation*}
        \mathrm{sinc}(x) \equiv \frac{\sin(\pi x)}{\pi x}.
    \end{equation*}

The detailed calculations are presented in \ref{appendix:calc_ft}. We note that Equation \ref{eq:analytical_ft_solution} contains the expected sinc terms necessary to reproduce the observed interference pattern in the $uv$-plane.

\subsection{Comparison with Simulation}\label{sec:comparison_with_simulation}

To validate the analytical model, we compare it against numerical simulations. Specifically, we simulate the observation containing the moving source (OBSID 1160763448) using the \texttt{pyuvsim} package \citep{Lanman2019}. As earlier, we first start by selecting down to the first time integration, which was found to contain the highest level of RFI. The second time integration is also retained for reasons discussed below.

\begin{figure*}
    \centering
    \includegraphics[width=1.0\linewidth]{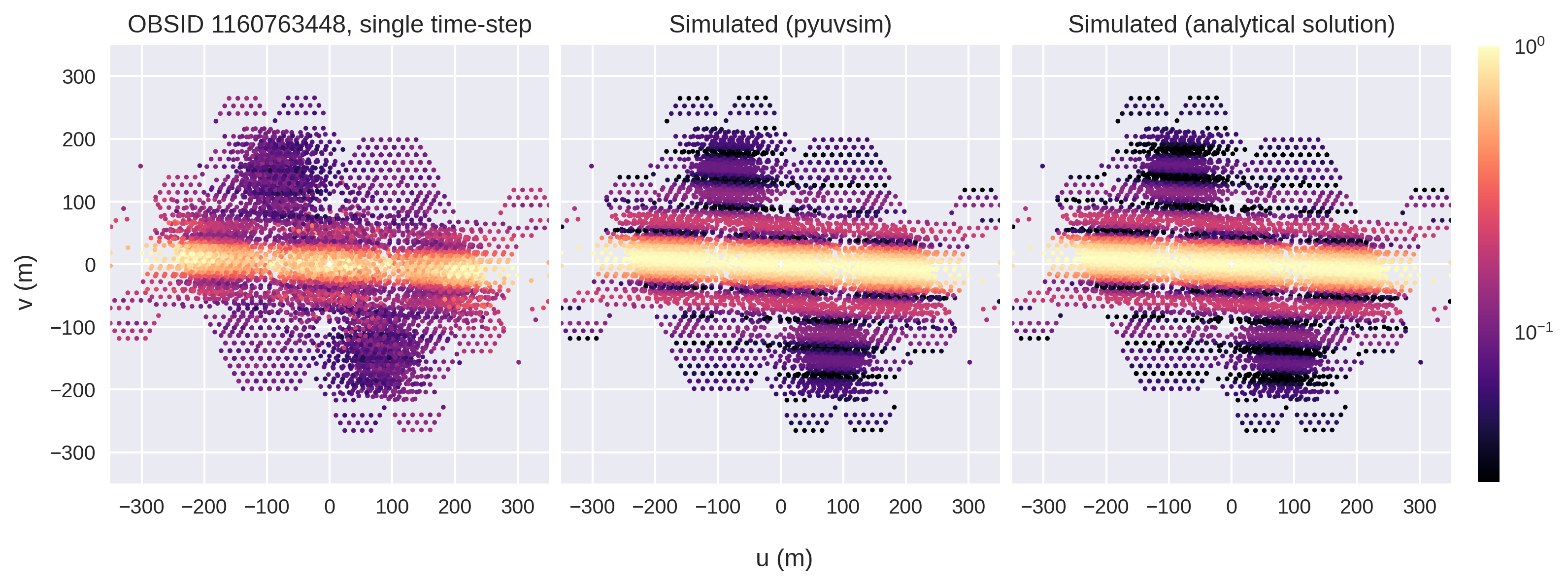}
    \caption{Using the right ascension and declination of the real RFI emitter found in OBSID 1160763448, it is possible to simulate the pattern it produces in the $uv$-plane. All plots show the peak-normalized maximum visibility amplitude per baseline at a single time integration, averaged across the DTV-7 band. Note that the colour is on a log scale. Left: First time integration of OBSID 1160763448. Centre: Simulation via \texttt{pyuvsim} (see Section \ref{sec:comparison_with_simulation}). Right: Simulation via the analytical Fourier transform from Equation \ref{eq:analytical_ft_solution}.}
    \label{fig:compare_data_sim_analytic}
\end{figure*}

Given these two adjacent integrations, we proceed as follows. Using \texttt{WSClean}, we image both integrations and extract the right ascension ($\alpha$) and declination ($\delta$) coordinates of the RFI emitter at its average position, yielding $(\alpha_1, \delta_1)$ for the first integration and $(\alpha_2, \delta_2)$ for the second. We then perform a linear interpolation between these two points to generate 200 evenly spaced coordinates $(\alpha_i, \delta_i)$, providing a smooth approximation to the source's motion. For each coordinate pair, we run an independent \texttt{pyuvsim} simulation with a single source at that location. The resulting 200 simulated datasets are then coherently averaged into a single two-second integration using the interferometric Python package \texttt{pyuvdata} \citep{pyuvdata, pyuvdata2}. This process preserves the sinc-like interference pattern in the $uv$-plane, which would otherwise be lost if we simulated only the endpoints $(\alpha_1, \delta_1)$ and $(\alpha_2, \delta_2)$. An example of the interference pattern produced using the output of \texttt{pyuvsim} is shown in the middle panel of Figure \ref{fig:compare_data_sim_analytic}.

Although Equation \ref{eq:analytical_ft_solution} is derived for exact $uv$ coordinates, real interferometers do not sample infinitely sharp points in the $uv$-plane. Finite antenna size smooths the response: each visibility corresponds to an average of the electric field across the physical collecting area. For the MWA, each tile spans approximately 5m $\times$ 5m \citep{Lonsdale_2009}. At the mean observing frequency of the DTV-7 band, 184.5 MHz, the wavelength is:
\begin{equation}
    \lambda = \frac{c}{\nu} = \frac{2.998\times 10^8 \text{m/s}}{184.5\times 10^6 \text{Hz}} \simeq 1.62 \text{m}
\end{equation}

Thus an MWA tile spans roughly:
\begin{equation}
    \frac{5 \text{m}}{1.62 \text{m}} \simeq 3.08
\end{equation}
wavelengths across. This sets a smoothing scale in the uv-plane; features evolving faster than this scale---such as deep, sharp nulls in the analytical solution---would be smoothed out.

Thus, in order to compare the simulated result to our analytical solution from Equation \ref{eq:analytical_ft_solution}, we start by initializing an evenly spaced grid of $u$ and $v$ values. The grid limits are determined by the physical maximum separation between antennas in the MWA Phase II layout. Having already obtained $(\alpha_1, \delta_1)$ and $(\alpha_2, \delta_2)$ in the previous step, we use Equation \ref{lm_ra_dec_conversion} to obtain the corresponding $\ell$ and $m$ coordinates. We then use simple arithmetic to convert $(\ell_0, m_0)$ and $(\ell_1, m_1)$ into a length $L$ and an angle $\theta$ in order to apply Equation \ref{eq:analytical_ft_solution} to our $uv$ grid. We use an arbitrarily small value for $W$, choosing $W = 10^{-4}$. This choice is motivated by numerical tests showing that the exact value of $W$ has negligible impact on the resulting comparison \footnote{As seen in Equation \ref{eq:analytical_ft_solution}, $W$ controls the width of the sinc-like pattern in the direction transverse to the main fringe pattern. In our analysis, this transverse structure is a subdominant effect, so $W$ does not require careful tuning.}.

This analytic solution is then convolved with a Gaussian kernel whose full width at half maximum (FWHM) is set to 3.08 wavelengths. This is done using the \texttt{scipy} Python package \citep{scipy}.

 Finally, the pattern is interpolated to the specific $uv$-coordinates corresponding to the real MWA layout. An example of the analytical pattern thus obtained is given in the right panel of Figure \ref{fig:compare_data_sim_analytic}. To match the observational and simulated scales, all plotted amplitudes are peak-normalized to their respective maxima.

We find excellent qualitative agreement between the real observation, the \texttt{pyuvsim} simulation, and the analytical solution. Additionally, we note that the analytical approach is several orders of magnitude faster to compute than the simulated approach. This efficiency gap arises because \texttt{pyuvsim} must resolve the event into many fine sub-integrations and account for the full MWA antenna configuration across the entire contaminated frequency range (181–188 MHz), both of which impose substantial computational demands. For instance, a single \texttt{pyuvsim} simulation can take upwards of an hour to run on 12 Intel Xeon Platinum CPU cores, whereas the analytical solution is obtained in seconds.

\section{Consequences for RFI flagging strategies} \label{sec:flagging}

Sections \ref{sec:stationary_vs_moving} and \ref{sec:modeling} demonstrated how and why moving RFI sources produce a distinct interference pattern in the $uv$-plane, leading to the suppression of contaminated signals on some baselines. Suppressed RFI is less likely to be picked up by per-baseline flagging algorithms, resulting in a flag mask that varies across baseline. A baseline-aggregated flagging strategy, on the other hand, can use the bright baselines to inform a global flag mask that would be equally applied to all baselines. The preferred flagging approach is not immediately obvious: unmodeled RFI and flagging can both introduce bias in a power spectrum analysis, albeit through different mechanisms. In this section, we attempt to disentangle these competing effects.

This section begins with examples of flagging masks in Section \ref{sec:flag_mask}, followed by the experimental design for our power spectrum tests in Section \ref{sec:experimental_design}. In Section \ref{sec:test1}, we examine the case of fixed RFI occupancy while varying the total integrated observing time. In Section \ref{sec:test2}, we instead fix the integration time and vary the RFI occupancy. We then discuss some limitations of these experiments in Section \ref{sec:limitations}. Finally, we benchmark against clean data in Section \ref{sec:benchmark}.

\subsection{Examples of flagging masks}\label{sec:flag_mask}

\begin{figure*}[ht]
    \centering
    \includegraphics[width=0.9\linewidth]{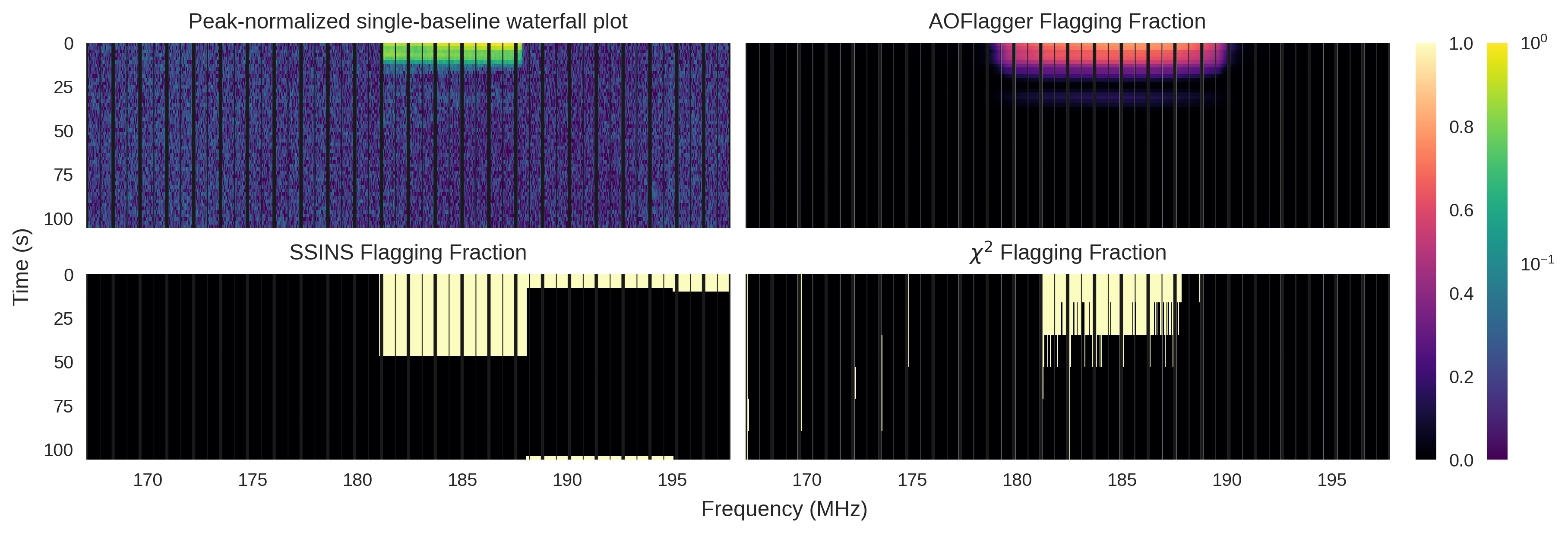}
    \caption{Top left: Single-baseline waterfall plot for OBSID 1160763448 after calibration and sky-subtraction, displayed on a logarithmic colour scale to highlight the RFI structure and peak-normalized so that both colour bars remain dimensionless. Top right: \texttt{AOFlagger} flagging fraction, calculated by averaging the output \texttt{AOFlagger} flag array along the baseline axis. Bottom left: \texttt{SSINS} flagging fraction. Bottom right: $\chi^2$ flagging fraction. The grey vertical bands correspond to the periodic MWA coarse-band edges, which are normally fully flagged (value 1) but are rendered in grey here for readability.}
    \label{fig:flag_comparison}
\end{figure*}

 To illustrate the difference between baseline-aggregated flagging and per-baseline flagging, we apply three different methods to OBSID 1160763448. We select flaggers that are commonly used with MWA data.

The first is \texttt{AOFlagger}, an iterative statistical flagger that operates independently on each baseline. \texttt{AOFlagger} estimates a smooth background model for the visibility amplitudes in time-frequency space, subtracts this model to obtain residuals, and then identifies samples or windows whose residuals exceed statistically derived thresholds. It then applies morphological operations (e.g. merging of adjacent time/frequency bins) to produce a cleaner, more coherent flag mask for each baseline. For our analysis, the standard settings for the MWA flagging strategy are applied \citep{AOFlagger}.

To compare \texttt{AOFlagger} to the other algorithms, we average its per-baseline flag mask along the baseline axis, producing a flagging rate array. This is plotted for OBSID 1160763448 in the top right panel of Figure \ref{fig:flag_comparison}. Notably, no single time-frequency bin achieves a 100\% \texttt{AOFlagger} flagging rate, in line with the above claim that per-baseline flaggers do not detect RFI on baselines where the averaging effect has suppressed the RFI the most.

Next, we apply the \texttt{SSINS} flagging algorithm, which first constructs per-baseline time-differenced visibilities to remove slowly varying sky signals and then incoherently averages the resulting amplitudes across baselines to form a single dynamic spectrum which, in the absence of RFI, should consist of Gaussian noise. Using the mean and variance estimated per frequency channel, each sample is converted to a z-score, and an iterative matched filter is applied to identify characteristic RFI signatures \citep{Wilensky_2019}. We apply the default significance threshold of 5$\sigma$. The \texttt{SSINS} flag mask for OBSID 1160763448 is shown in the bottom left panel of Figure \ref{fig:flag_comparison}.

Finally, we apply the $\chi^2$ flagging algorithm, which relies on redundant calibration. In a redundant array, baselines with identical geometric lengths and orientations should, in principle, measure identical visibilities. Redundant calibration exploits this redundancy to solve simultaneously for complex antenna gains and a set of unique visibilities per redundant group. After calibration, the algorithm computes a per-time, per-frequency, per-polarization $\chi^2$ statistic that quantifies how strongly measured visibilities deviate from the calibrated redundant model. Anomalously high $\chi^2$ values (compared to the expected thermal noise threshold) are indicative of potential RFI events, and these samples are flagged. This produces one flag array per polarization \citep{Kunicki_Pober_2024}. For consistency with \texttt{SSINS}, which outputs a single global flag mask, we take the union of the $\chi^2$ flags across polarizations. The output of $\chi^2$ is shown for OBSID 1160763448 in the bottom right panel of Figure \ref{fig:flag_comparison}. As noted in \cite{Kunicki_Pober_2024}, the $\chi^2$ values from redundant calibration are useful for identifying RFI, but the algorithm that produces flags from those values is still likely sub-optimal, especially in relation to the MWA coarse band lines.  As such, the vertical streaks at low-frequencies are almost certainly false positives and should not distract from the main point: the $\chi^2$ approach is to create a global flag mask that applies to all baselines, regardless of whether there is detectable RFI on any particular baseline.

While the \texttt{AOFlagger} approach does not flag the suppressed contamination present on some baselines, and may therefore leave faint RFI in the data, it does retain a larger total number of $uv$-modes by including these weakly contaminated baselines. On the other hand, \texttt{SSINS} and $\chi^2$ remove a larger fraction of the contamination, but at the cost of reduced $uv$-coverage. This creates a direct trade-off between contamination removal and data retention.

The analytical treatment developed in Section \ref{sec:analytical_model} allows us to approximate the level of RFI suppression in the nulls. For the event considered here, the peak-to-null ratio of the analytic pattern (i.e. the peak-to-null ratio of the data from the right panel of Figure \ref{fig:compare_data_sim_analytic}) was found to be approximately 750. That is, destructive interference reduces the RFI amplitude by only about three orders of magnitude relative to the peaks. Considering that RFI events are typically many orders of magnitude brighter than the cosmological EoR signal (often $\geq 10^5-10^7$ times brighter), a suppression factor of only $\sim 10^3$ means that residual RFI not flagged by a per-baseline algorithm remains brighter than the EoR, and therefore relevant in the context of our analysis.

\subsection{Experimental design}\label{sec:experimental_design}

The results of the previous section show that different flagging strategies make different compromises between residual contamination and loss of $uv$-coverage. To determine which compromise is most relevant for EoR science, we now test how these strategies affect the recovered power spectrum. The goal is to determine which method produces the least biased power spectrum.

We construct a controlled set of power spectrum experiments by combining real MWA data with simulated RFI injections. We begin by identifying a sample of 300 clean 2-minute MWA Phase II observations from the fall 2016 observing season. These observations are selected by applying several flagging algorithms, including $\chi^2$ and \texttt{AOFlagger}, and requiring that no flags are returned. This clean sample forms the basis for a realistic, deep power spectrum with sufficient $uv$-coverage. 

We then inject a simulated RFI event into a subset of these observations. The simulated event is the 12-second RFI source generated with \texttt{pyuvsim} and described in Section \ref{sec:comparison_with_simulation}. The injection is performed at the waterfall level: for each observation in the subset, we iterate over baselines and add the simulated RFI event directly to the corresponding visibilities. This approach gives direct control over the brightness and timing of the RFI. 

Although the same simulated event is reused, it does not add coherently across observations. The source trajectory is fixed in local horizontal coordinates (altitude and azimuth), but each observation is taken at a different time, so the corresponding right ascension coordinates change with Earth's rotation. This results in the apparent trajectory tracing different paths in celestial coordinates for each observation, rather than overlapping each other and accumulating as a single brighter event.

A subset is randomly selected from the 30 clean observations and copied to create a separate injection sample, leaving the original clean observations unchanged. Each copied observation receives one simulated RFI event. The injected flux density is drawn uniformly between 1 and 10 kJy, spanning the brighter end of the flux density distribution measured from real RFI events in the unused remainder of the 2016 observing season. The event is restricted to the Australian DTV-7 frequency band in all cases, while its start time is randomly selected independently for each injected observation.

For each injected observation, we construct four flagging treatments:

  \begin{itemize}
      \item[1.] \texttt{unflagged}: the injected RFI is left unflagged.
      \item[2.] \texttt{flagged}: the contaminated time-frequency region is flagged on all baselines.
      \item[3.] \texttt{all\-freq\-flagged}: all frequencies are flagged at the contaminated time-steps on all baselines.
      \item[4.] \texttt{aof}: the per-baseline \texttt{AOFlagger} algorithm is applied.
  \end{itemize} 

The \texttt{flagged} treatment represents an idealized baseline-aggregated flagging strategy, similar in spirit to methods such as \texttt{SSINS} or $\chi^2$. Because the RFI is simulated, the contaminated time-steps and frequency channels are known exactly, allowing us to flag the affected region without relying on a detection threshold.

The \texttt{all\-freq\-flagged} treatment is motivated by the findings of \cite{wilensky_2022}, who showed that chromatic RFI flags can produce excess power in the EoR window in a manner similar to residual RFI. Flagging the entire analysis frequency band when RFI is detected therefore provides a way to avoid introducing chromatic flagging structure.

All data sets, including the 300 clean observations and the four flagging variants of the RFI-injected copied observations, are processed through the FHD/$\varepsilon$ppsilon calibration and power spectrum estimation pipeline \citep{Barry_2019}. We then compare the resulting power spectra against the corresponding clean reference spectra. 

To isolate the effects of flagging, we eliminate two additional sources of bias. First, to avoid calibration bias from the injected RFI, we use the FHD/$\varepsilon$ppsilon calibration solutions obtained from the clean observations and apply those same solutions to the corresponding RFI-injected data sets. Second, we restrict the analysis to the ``dirty'' power spectra, computed directly from the calibrated visibilities. We do not use ``residual'' power spectra, which are formed after subtracting a sky model from the calibrated data, because the model itself depends on the flagging strategy and would introduce an additional confounding factor.

Since we are specifically interested in the effects of flagging for EoR applications, we restrict the $k$-modes entering our analysis to those located above the foreground wedge. When looking at one-dimensional power spectra, we further restrict $k$ values to $0.1 < k < 0.5 \, h\mathrm{Mpc}^{-1}$, corresponding to the modes with the highest signal-to-noise ratio within the EoR window. Finally, we also remove the $k$ modes associated with the MWA's first coarse-band edge.

\subsection{Fixed RFI occupancy with varying total integrated observing time}\label{sec:test1}

We first isolate the effect of a single RFI event by injecting one event into data sets with different number of total integrated observations. Specifically, we combine one contaminated observation with otherwise clean samples of 10, 100, and 300 observations, corresponding to approximately 0.3, 3, and 10 hours of data. For each case, we compare the resulting integrated power spectrum to that obtained from the same observations wthout the RFI injection. This experiment tests how the impact of a fixed RFI event changes as it is diluted into increasingly deep integrations.

We perform this experiment for both a 1-kJy and 10-kJy injected event. For the fainter 1-kJy case, we find that the bias introduced by flagging can be larger than the bias introduced by leaving the RFI unflagged. In this regime, the loss of $uv$-coverage and the spectral structure introduced by flagging dominate over the relatively weak residual contamination. 

\begin{figure*}
    \centering
    \includegraphics[width=0.95\linewidth]{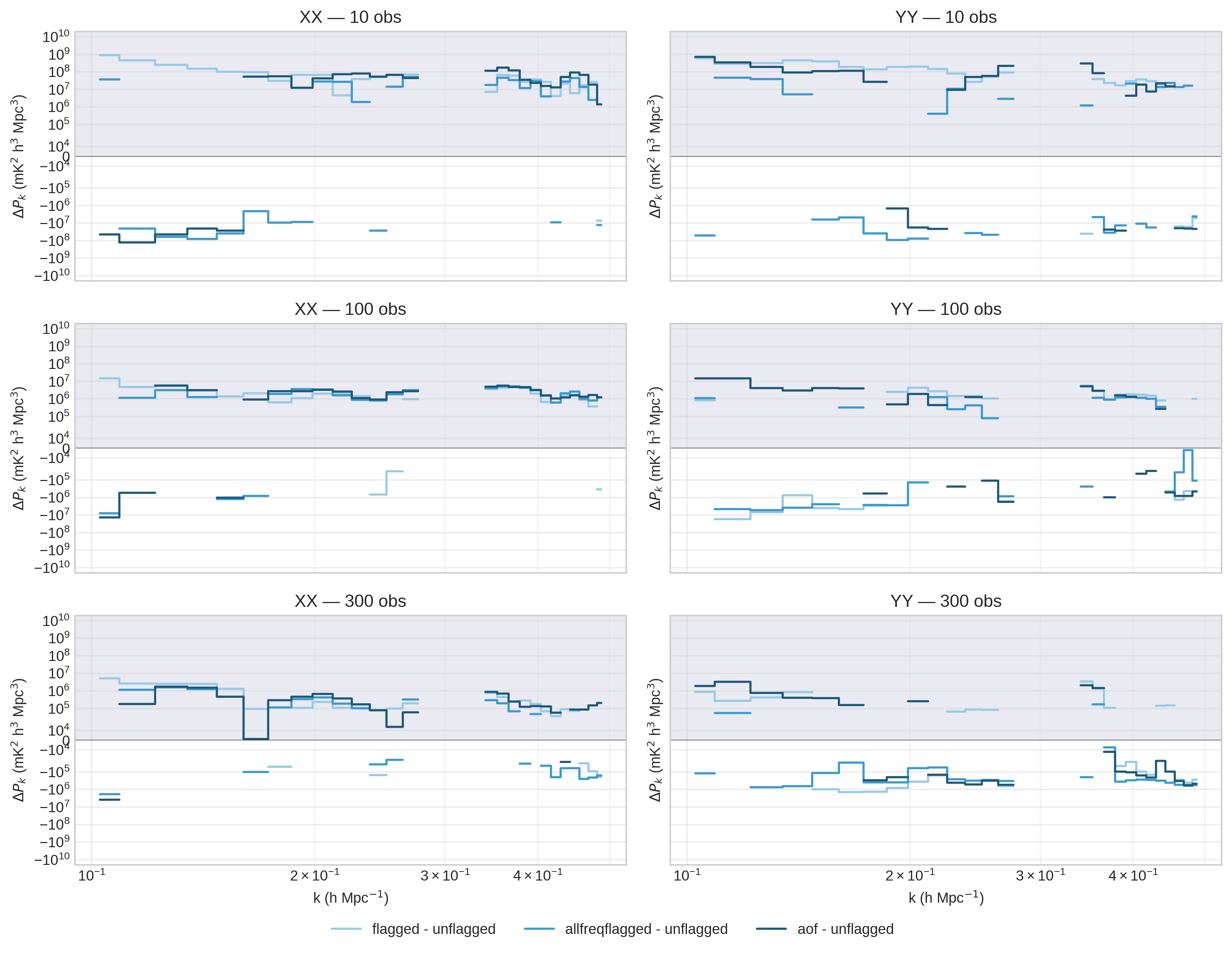}
    \caption{1D power spectrum differences between each flagging treatment and the corresponding \texttt{unflagged} case for a 1-kJy injected RFI event. The columns show the XX and YY polarizations. The rows correspond to power spectra formed from 10, 100, and 300 total observations, equivalent to approximately 0.3, 3, and 10 hours of data. Positive values indicate modes where the flagging treatment produces more excess power than the \texttt{unflagged} case. All flagging treatments produce excess power compared to the \texttt{unflagged} case.}
    \label{fig:1ky_flagging_comparison}
\end{figure*}

This behavior is shown in Figure \ref{fig:1ky_flagging_comparison}, which presents differences in the spherically averaged one-dimensional power spectrum, $P_k$, hereafter referred to as the 1D power spectrum, between each flagging treatment and the corresponding \texttt{unflagged} case for 10, 100, and 300 total integrated observations. Specifically, we subtract the flagging-treatment power spectrum from the \texttt{unflagged} power spectrum. Positive values indicate modes where the flagging treatments are producing more excess power than contamination from unflagged RFI. In all cases, we notice an obvious positive bias.

This result is somewhat surprising. Indeed, \cite{wilensky_2022} have applied the \texttt{all\-freq\-flagged} method to simulated RFI events using the MWA Phase I layout and found that the increase in the power spectrum compared to a clean reference was sub-EoR. The discrepancy with our results may point directly to a weakness in redundant array laouts, such as the MWA Phase II (compact) layout considered here. Indeed, the Phase II layout contains many less total $uv$-modes than the Phase I layout, meaning that any addition loss of $uv$-coverage leaves a much bigger footprint in the power spectrum. The discrepancy may also be explained by the findings of \cite{Gehlot_2024}, who show that transient RFI decorrelates faster than static sources, potentially changing how strongly it contributes to the resulting power spectrum.

\begin{figure*}
    \centering
    \includegraphics[width=0.95\linewidth]{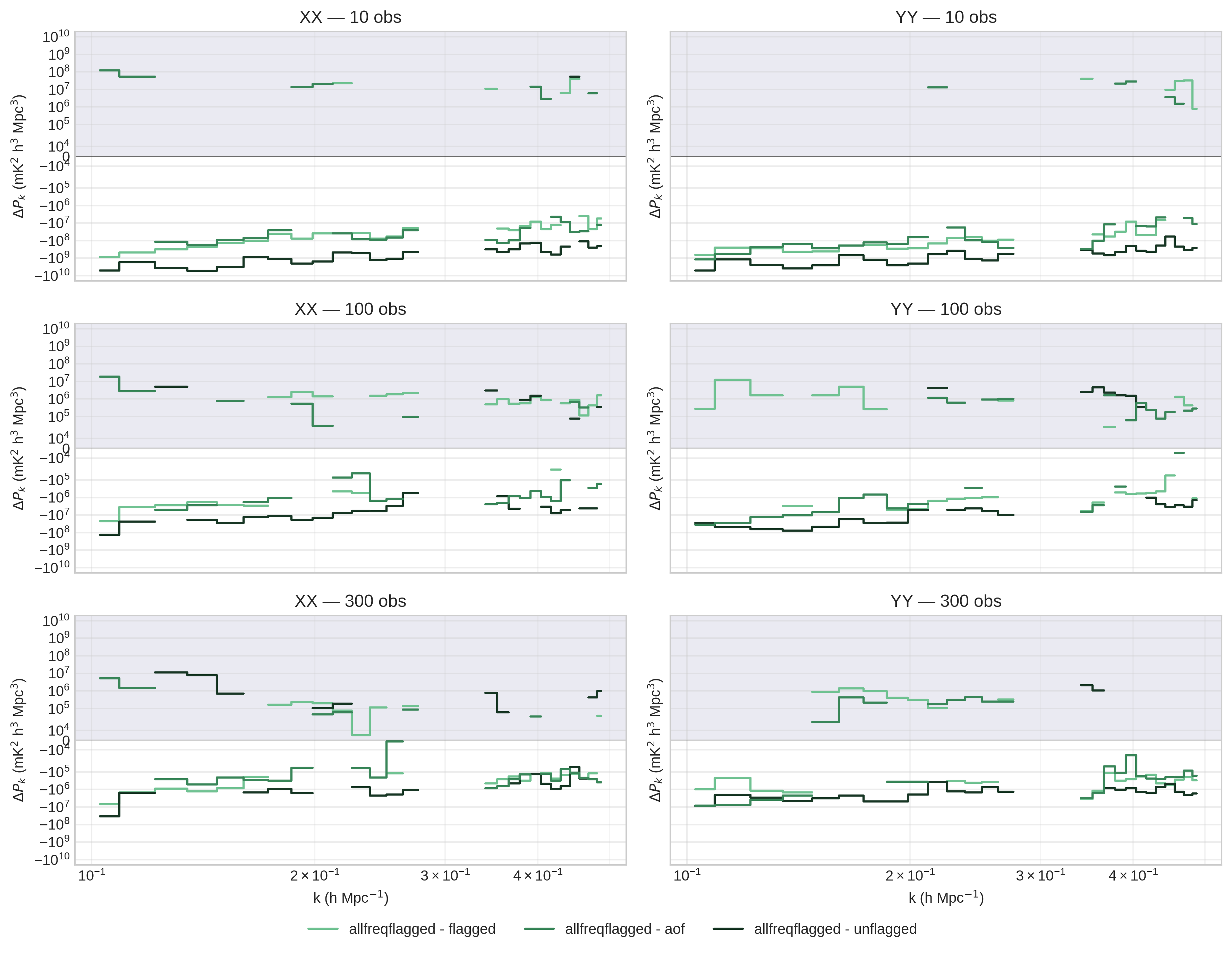}
    \caption{1D power spectrum differences between the \texttt{all\-freq\-flagged} treatment and all remaining flagging treatments for a 10-kJy injected RFI event. The columns show the XX and YY polarizations. The rows correspond to power spectra formed from 10, 100, and 300 total observations, equivalent to approximately 0.3, 3, and 10 hours of data. Negative values indicate modes where \texttt{all\-freq\-flagged} produces less excess power than the the other case. While \texttt{all\-freq\-flagged} is strongly preferred for the deepest integration, the preference becomes less marked as the total number of integrated observations increases.}
    \label{fig:10kjy_allfreqflagged_minus_rest}
\end{figure*}

Next, we repeat the experiment with a bright 10-kJy event. In this regime, leaving the RFI unflagged yields more excess power than any of the flagging treatments, making \textit{some} form of flagging clearly preferable. It is therefore useful to compare the flagging treatments directly. In keeping with the findings of \cite{wilensky_2022}, we use \texttt{all\-freq\-flagged} as the reference treatment and compute 1D power spectrum differences between \texttt{all\-freq\-flagged} and each of the other flagging approaches. These comparisons are shown in Figure \ref{fig:10kjy_allfreqflagged_minus_rest}. For this set of difference plots, negative values indicate modes where \texttt{all\-freq\-flagged} produces less excess power than the comparison treatment, while positive values indicate modes where it performs worse.

 The clearest preference for \texttt{all\-freq\-flagged} occurs in the shallowest data set, containing 10 total observations. As the total integration depth increases, this preference becomes less pronounced. By 300 observations, corresponding to approximately 10h of data, the single 10-kJy event is sufficiently diluted that \texttt{all\-freq\-flagged} is no longer clearly distinguishable from the other flagging treatments.

This first experiment establishes that, in the limit of faint RFI, leaving the RFI unflagged can introduce less power spectrum bias than applying a flagging treatment. As the RFI brightness increases, residual contamination becomes more important, and flagging becomes the preferred option. For bright isolated events, the \texttt{all\-freq\-flagged} performs best for shallow integrations, while deeper integrations are more or less inconclusive.

\subsection{Fixed total integrated observing time with varying RFI occupancy}\label{sec:test2}

We perform a complementary experiment in which the total observing depth is fixed while the number of RFI events is varied. Here, we specifically examine the regime in which the loss of individual $uv$-samples is less important due to dilution in a large number observations, while residual contamination can accumulate across multiple events. In all cases, the power spectrum is formed from a set of 300 integrated observations, corresponding to approximately 10 hours of data.  We construct two contaminated samples by randomly selecting either 5 or 30 observations from the injection set and combining them with enough clean observations to reach a total of 300. The resulting integrated power spectra are then compared across the different flagging treatments. The flux density of each injected RFI event is drawn uniformly between 1 and 10 kJy, following the procedure described in Section \ref{sec:experimental_design}.

\begin{figure*}
    \centering
    \includegraphics[width=0.95\linewidth]{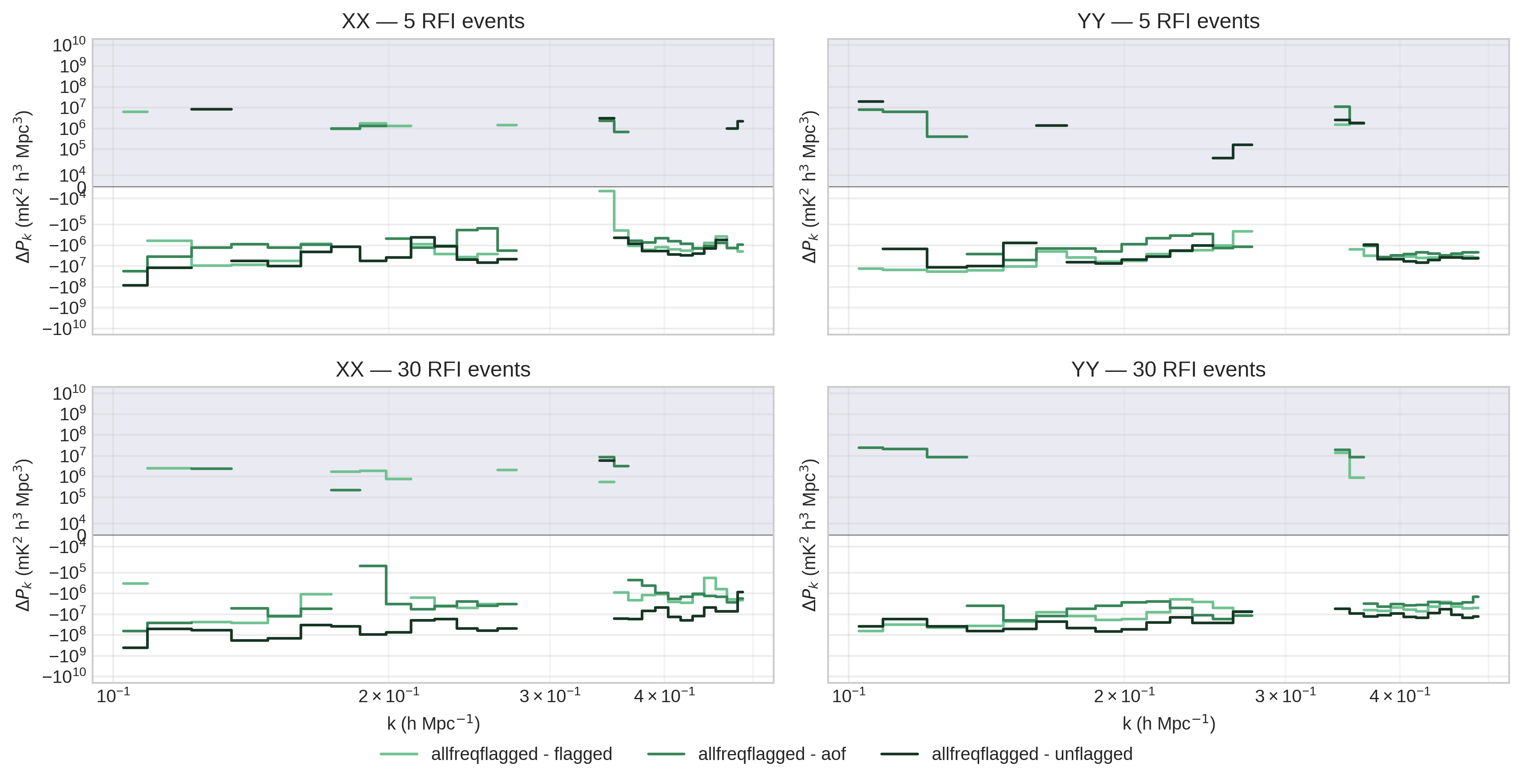}
    \caption{1D power spectrum differences between \texttt{all\-freq\-flagged} and the other flagging treatments for fixed 300-observation integrations. The columns show the XX and YY polarizations. The rows correspond to cases with 5 and 30 injected RFI events. Negative values indicate modes where \texttt{all\-freq\-flagged} produces less excess power than the comparison treatment. While \texttt{all\-freq\-flagged} is already preferred for 5 injected events, the preference becomes more pronounced when the number of injected events is increased to 30.}
    \label{fig:N_allfreqflagged_minus_N_rest}
\end{figure*}

This experiment strengthens the case for the \texttt{all\-freq\-flagged} strategy. As the number of RFI events increases, the cumulative impact of residual RFI becomes more important, and \texttt{all\-freq\-flagged} more consistently produces less excess power than the other treatments. This is supported by Figure \ref{fig:N_allfreqflagged_minus_N_rest}, which presents 1D power spectrum differences between \texttt{all\-freq\-flagged} and the other flagging treatments for 5 and 30 injected RFI events.

\subsection{Experiment limitations}\label{sec:limitations}

Within the scope of these experiments, it remains difficult to distinguish cleanly between the \texttt{flagged} and \texttt{aof} methods. In both cases, the interpretation is complicated by several competing contributions to the recovered power spectrum: residual contamination left in the data, loss of $uv$-coverage, and power leakage introduced by the chromatic structure of the flag mask. The \texttt{flagged} treatment removes the known contaminated time-frequency region on all baselines, but does so at the cost of reduced $uv$-coverage. The \texttt{aof} treatment preserves more $uv$-modes by flagging on a per-baseline basis, but can leave faint residual RFI on baselines where the moving source signal is suppressed below the detection threshold. Meanwhile, both treatments produce chromatic flags leading to leakage in the EoR window.

\begin{figure*}
    \centering
    \includegraphics[width=0.95\linewidth]{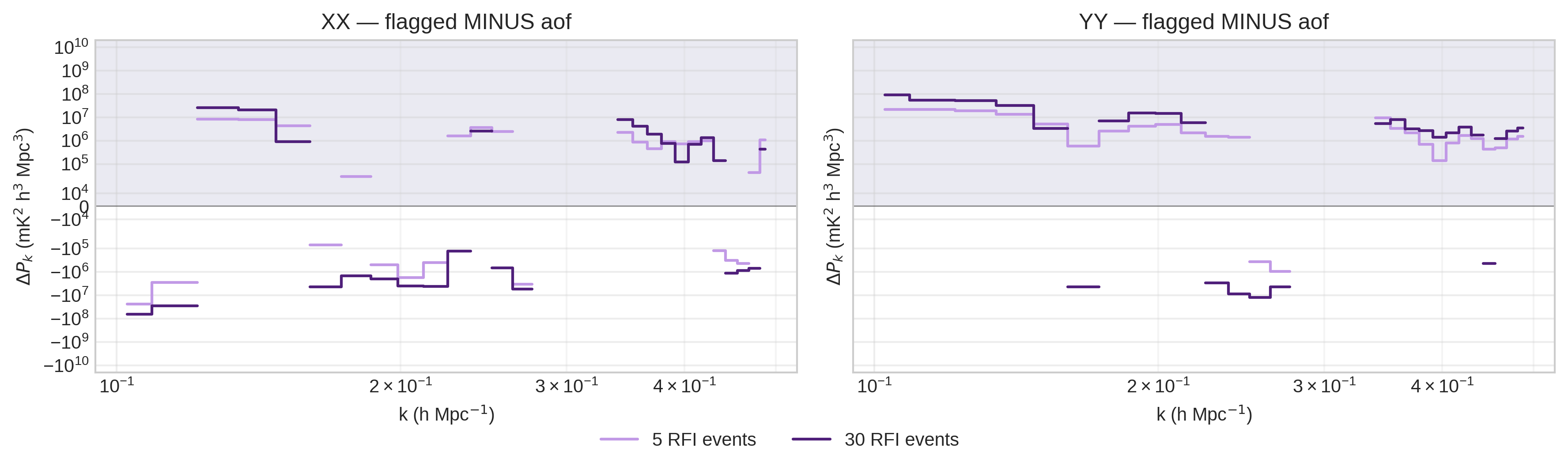}
    \caption{1D power spectrum differences between \texttt{flagged} and \texttt{aof} for fixed 300-observation integrations. The columns show the XX and YY polarizations. The rows correspond to cases with 5 and 30 injected RFI events. Negative values indicate modes where \texttt{aof} produces less excess power than the comparison treatment. Neither treatment performs consistently better across $k$-modes and polarizations, so these results do not clearly favour one method over the other.}
    \label{fig:N_flagged_minus_N_aof}
\end{figure*}

Even after focusing on deeper 10-hour integrations to reduce the importance of individual missing $uv$-samples, and after increasing the number of injected RFI events to amplify the effect of residual contamination, these contributions remain difficult to separate. This is shown in Figure \ref{fig:N_flagged_minus_N_aof}, which directly compares the \texttt{flagged} and \texttt{aof} treatments for 5 and 30 injected RFI events. The differences do not reveal a consistent preference for either treatment across polarization, $k$-mode, and RFI occupancy. Indeed, within the parameter space tested here, the two approaches appear to produce comparable levels of power spectrum bias. We therefore do not make a conclusive claim about whether \texttt{flagged} or \texttt{aof} should be preferred for 21-cm power spectrum analyses.

\subsection{Benchmarking against uncontaminated data}\label{sec:benchmark}

As a final comparison, we directly compare each RFI-injected data set to a corresponding uncontaminated reference. In the previous experiments, we primarily compared flagging treatments against one another. While this identifies which treatment performs best among the tested options, it does not show how much bias remains relative to a truly clean data set.

\begin{figure*}
    \centering
    \includegraphics[width=0.95\linewidth]{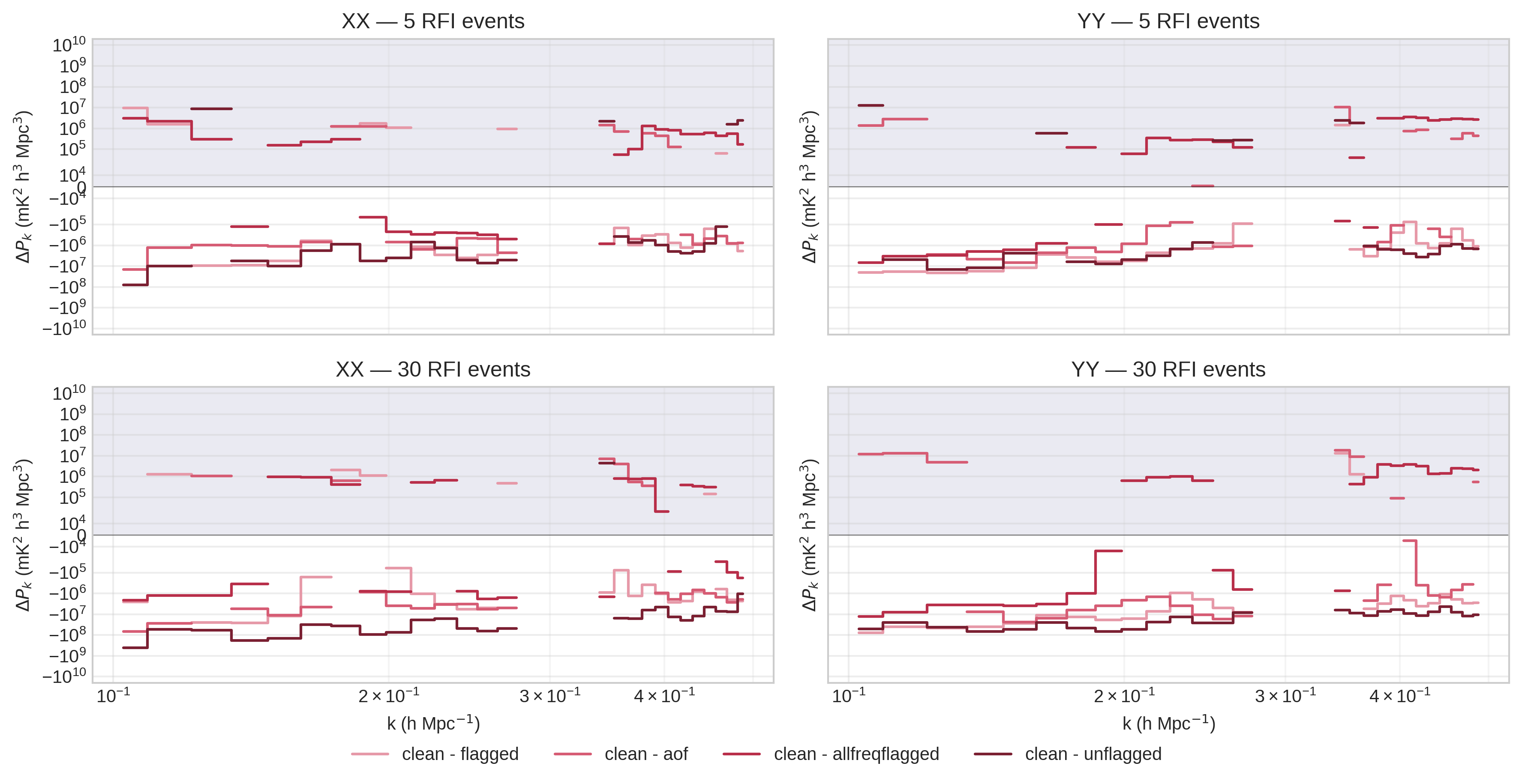}
    \caption{1D power spectrum differences between the clean 300-observation reference spectrum and RFI-injected spectra processed with each flagging treatment. The columns rows correspond to cases with 5 and 30 injected RFI events. Negative values indicate modes where the RFI-injected treatment produces more excess power than the clean reference. Across the tested cases, the clean reference remains less biased than the spectra produced by any of the flagging treatments.}
    \label{fig:clean_comparison}
\end{figure*}

All of the experiments described above were benchmarked against their corresponding clean reference spectra. We illustrate this comparison here with a representative case. We compare the 300-observation clean power spectrum to the RFI-injected power spectra produced under each treatment: \texttt{unflagged}, \texttt{flagged}, \texttt{all\-freq\-flagged}, and \texttt{aof}. We perform this comparison for both 5 and 30 injected RFI events. The result is shown in Figure \ref{fig:clean_comparison}. 

Across all experiments, the clean reference spectra consistently show less excess power than any of the RFI-injected cases. This is true even for the best-performing flagging treatment, \texttt{all\-freq\-flagged}. This result emphasizes that flagging can reduce the impact of RFI, but it does not fully restore the statistical properties of an uncontaminated data set. Residual contamination, lost $uv$-coverage, and chromatic flagging structure all leave measurable signatures in the recovered power spectrum.

\section{Discussion} \label{sec:discussion}

Although our analysis uses the MWA as a specific case study, many of the qualitative trade-offs identified here are expected to apply more broadly to low-frequency 21-cm power spectrum experiments. However, the quantitative conclusions should be interpreted within the scope of the array configuration and simulated RFI environment considered here.

Indeed, the conclusion that no flagging is preferred for relatively faint RFI events may be specifically related to the MWA's redundant Phase II layout. Redundant layouts have less $uv$-coverage by design, so it is natural that any additional loss of $uv$-modes results in more pronounced bias in the power spectrum. Instruments who use redundant layouts, such as CHIME or the upcoming CHORD experiment \citep{chord}, may be susceptible to the same loss in $uv$-coverage, while other instruments operating under different array configurations, such as LOFAR, may reach different conclusions.

Additionally, the experiments focus on the MWA's observing environment, using repeated injections of a representative DTV-7 event with flux densities between 1 and 10 kJy and occurrence rates typical of the 2016 MWA observing season. Instruments operating in substantially different RFI environments, where contamination is more frequent or exhibits different temporal or spectral characteristics (e.g., LOFAR), may exhibit qualitatively different behavior as the balance between residual contamination and flagging-induced bias changes.

Nevertheless, the MWA remains a relevant test-case, particularly because it is located in the same radio-quiet zone as the upcoming low-frequency telescope of the Square Kilometre Array Observatory (SKA-low), making our results directly informative for future observations at that site.

\section{Conclusion} \label{sec:conclusion}
In this work, we have developed an analytical framework for forward-modeling RFI from moving sources and validated its qualitative behaviour with real data and \texttt{pyuvsim} simulations. We showed that a moving source produces a sinc-like pattern in the $uv$-plane as its signal is averaged over a finite correlator integration. This causes the RFI amplitude to be strongly baseline-dependent, with the signal appearing substantially suppressed on some baselines.

This behavior has direct consequences for RFI flagging. Because per-baseline flaggers operate independently on each baseline, they can fail to identify moving-source contamination on baselines where the signal is suppressed below the detection threshold. Baseline-aggregated approaches are invulnerable to this specific failure mode, but they remove more data and can introduce additional bias through reduced $uv$-coverage and chromatic flagging structure.

To assess the impact of these effects on EoR power spectrum analyses, we constructed a set of controlled injection experiments using real MWA observations combined with simulated moving-source RFI events. These experiments allowed us to compare several treatments: leaving the RFI unflagged, flagging only the contaminated time-frequency region on all baselines, flagging all frequencies at contaminated time-steps, and applying \texttt{AOFlagger} on a per-baseline basis.

We find that the preferred treatment depends on the brightness and occupancy of the RFI. For faint isolated events, the bias introduced by flagging can exceed the bias from leaving the weak residual contamination unflagged. For brighter events, however, flagging becomes clearly preferable. Across the brighter and higher-occupancy cases tested here, the \texttt{all\-freq\-flagged} treatment most consistently produces the lowest excess power, in agreement with the expectation from \cite{wilensky_2022} that avoiding chromatic flagging structure is important for EoR power spectrum analyses.

The comparison between the \texttt{flagged} and \texttt{aof} treatments is less conclusive. The \texttt{flagged} treatment removes the known contaminated region on all baselines, but at the cost of reduced $uv$-coverage. The \texttt{aof} treatment preserves more $uv$-modes, but can leave faint residual contamination on baselines where the moving-source signal is suppressed. Both treatments also introduce chromatic flagging structure. Within the parameter space tested here, these competing effects produce comparable levels of power spectrum bias, and we therefore do not make a conclusive claim about which of the two should be preferred for EoR analyses. A more direct way to disentangle these contributions would be to construct a theoretical power spectrum estimate from contaminated and flagged data directly from first principles, following an approach similar to \cite{Wilensky_2020}. Such a calculation may make it possible to separate the effects of residual RFI, missing $uv$-coverage, and chromatic flagging structure, but lies beyond the scope of the present work.

Finally, as expected, our comparison against the fully clean reference spectra shows that none of the tested flagging treatments match up to the uncontaminated result. Even when \texttt{all\-freq\-flagged} performs best among the available strategies, the clean data remain less biased. This highlights a fundamental limitation of flagging-based mitigation: flagging can reduce RFI contamination, but it cannot recover the information lost through data excision, even when the flagging strategy is close to optimal.

It is possible that inpainting (i.e. interpolating missing or  flagged values) would help mitigate the observed adverse effects of flagging on the power spectrum. This, however, lies beyond the scope of the present work. The interested reader is encouraged to consult \cite{chen2025}.

All in all, these results motivate a shift toward RFI modeling and subtraction, rather than flagging alone. The analytical framework developed here provides an important step in that direction by describing the expected $uv$-space structure of moving-source RFI directly. Future work could use this model to fit and subtract moving RFI sources from the visibilities, with the goal of preserving $uv$-coverage while reducing residual contamination.

\begin{acknowledgement}
This work makes use of data obtained from Inyarrimanha Ilgari Bundara, the Murchison Radio-astronomy Observatory. We acknowledge the Wajarri Yamaji People as the Traditional Owners and native title holders of the land on which the Observatory is located. The establishment of CSIRO's Murchison Radio-astronomy Observatory was an initiative of the Australian Government, supported by the Government of Western Australia and the Science and Industry Endowment Fund. Operation of the Murchison Widefield Array is supported by the Australian Government through the National Collaborative Research Infrastructure Strategy (NCRIS), under a contract to Curtin University administered by Astronomy Australia Limited. This research was enabled by resources provided by the Pawsey Supercomputing Research Centre, funded by the Australian Government and the Government of Western Australia.

We would furthermore like to thank the anonymous referee whose comments on the first submission of this paper have allowed us to significantly expand its scope. We also thank Miguel Morales for his helpful suggestions that improved the presentation of the 1D power spectrum difference plots.

\end{acknowledgement}

\paragraph{Funding Statement}

This research was supported by a grant from the US National Science Foundation (award number 2228989) including a Graduate Research Supplement from the Spectrum Innovation Initiative.

\paragraph{Competing Interests}

None

\paragraph{Data Availability Statement}

The MWA data used in this work are available for download
at ASVO, \\https://asvo.mwatelescope.org/.

\printendnotes

\bibliography{main}

\appendix

\section{Analytical Fourier Transform of a Streak} \label{appendix:calc_ft}

In this section, we present the detailed calculations for the two-dimensional Fourier transform of the streak presented in Equation \ref{eq:analytical_ft_solution}.

We begin with our equation for a streak in image space, Equation \ref{eq:streak_eqn}.

\begin{equation*}
        f(\ell,m)=
        \begin{cases}
        1, & \left| s \right| \le \frac{L}{2},\quad \left| t \right| \le \frac{W}{2},\\
        0, & \text{otherwise,}
        \end{cases}
    \end{equation*}
    with
    \begin{equation*}
    \begin{split}
        s &= (\ell-\ell_0)\cos\theta + (m-m_0)\sin\theta,\\
        t &= -(\ell-\ell_0)\sin\theta + (m-m_0)\cos\theta.
    \end{split}
    \end{equation*}

We can invert the above to express $(\ell, m)$ in terms of $(s, t)$:

\begin{equation*}
    \begin{split}
        \ell &= \ell_0 + s \cos\theta - t \sin\theta, \\
  m    &= m_0 + s\sin\theta + t \cos\theta.
    \end{split}
\end{equation*}

The transformation from $(\ell,m)$ to $(s,t)$ is a rotation plus a
translation, so the Jacobian is unity:
\begin{equation*}
   J = \frac{\partial \ell}{\partial s} \frac{\partial m}{\partial t} - \frac{\partial \ell}{\partial t} \frac{\partial m}{\partial s} = 1.
\end{equation*}

Next, we use the definition of a two-dimensional Fourier transform, which was presented in Equation \ref{eq:fourier_2d}. Changing variables to $(s,t)$:

\begin{equation*}
  F(u, v)=\int_{-\infty}^{\infty} \int_{-\infty}^{\infty} f(s, t) e^{-j 2 \pi \bigl( u \ell(s, t) +v m(s, t) \bigr)} d s \, d t 
\end{equation*}

Because $f(s,t)=1$ only for $|s| \leq L/2$ and
$|t|\leq W/2$, the limits of integration become:
\begin{equation*}
  F(u,v) = \int_{s=-L/2}^{L/2} \int_{t=-W/2}^{W/2}
    e^{-i2\pi\bigl(u\ell(s,t) + v m(s,t)\bigr)}\, ds \, dt 
\end{equation*}

Substituting the expressions for $\ell$ and $m$,

\begin{equation*}
    \begin{split}
      u\ell + v m
        &= u\bigl(\ell_0 + s\cos\theta - t\sin\theta\bigr)
           + v\bigl(m_0 + s\sin\theta + t\cos\theta\bigr) \\
        &= u\ell_0 + v m_0
         + s\bigl(u\cos\theta + v\sin\theta\bigr)
         + t\bigl(-u\sin\theta + v\cos\theta\bigr)
    \end{split}
\end{equation*}

Adopting the notation from Equation \ref{eq:analytical_ft_solution}, we define:

\begin{equation*}
    \begin{split}
      u_\parallel &= u\cos\theta + v\sin\theta\\
      u_\perp     &= -u\sin\theta + v\cos\theta
    \end{split}
\end{equation*}

Then:

\begin{equation*}
  u\ell + v m = u\ell_0 + v m_0 + s\,u_\parallel + t\,u_\perp 
\end{equation*}

Thus the Fourier transform factorises as:

\begin{equation*}
    \begin{split}
      F(u,v) 
        &= \int_{-L/2}^{L/2} \int_{-W/2}^{W/2}
           \exp\!\left[-i2\pi\bigl(u\ell_0 + v m_0 + s\,u_\parallel + t\,u_\perp\bigr)\right]
           \, dt\,ds \\
        &= e^{-i2\pi(u\ell_0 + v m_0)}
           \left[ \int_{-L/2}^{L/2} e^{-i2\pi u_\parallel s}\,ds \right]
           \left[ \int_{-W/2}^{W/2} e^{-i2\pi u_\perp t}\,dt \right]
    \end{split}
\end{equation*}

We can evaluate each piece separately. For $u_\parallel$,

\begin{equation*}
    \begin{split}
      \int_{-L/2}^{L/2} e^{-i2\pi u_\parallel s}\,ds
        &= \left.\frac{i e^{-i2\pi u_\parallel s}}{2\pi u_\parallel}\right|_{-L/2}^{L/2} \\
        &= \frac{\sin(\pi u_\parallel L)}{\pi u_\parallel} \\
        &= L \, \mathrm{sinc}(u_\parallel L),
    \end{split}
\end{equation*}

where we adopt the normalized sinc function, $\mathrm{sinc}(x) = \sin(\pi x)/\pi x$.

The calculation is identical for $u_\perp$, and we obtain:

\begin{equation*}
    \begin{split}
      \int_{-W/2}^{W/2} e^{-i2\pi u_\perp t}\,dt
        &= W \mathrm{sinc}(u_\perp W)
    \end{split}
\end{equation*}

Combining the pieces:
\begin{equation*}
\begin{split}
  F(u,v) = L \, W
        \operatorname{sinc}(L  u_\parallel)\,
        \operatorname{sinc}(W u_\perp) \, e^{-i2\pi(u\ell_0+vm_0)} \\ 
\end{split}
\end{equation*}

This is Equation \ref{eq:analytical_ft_solution}.

\section{More examples} \label{appendix:uv_plane}

This appendix presents additional examples of the baseline-dependent visibility-amplitude patterns produced by moving sources of RFI, using MWA observations. In each figure, the left panel shows a calibrated, sky-subtracted waterfall plot for a single baseline, while the right panel shows the per-baseline peak-normalized visibility amplitudes averaged over the time-frequency region outlined by the red rectangle. Each rectangle spans 2 seconds, beginning at the time specified in the corresponding caption, and covers either the DTV-7 band (181-188 MHz) or the approximate emission range identified from the waterfall plot.

\begin{figure*}
    \centering
    \includegraphics[width=0.9\linewidth]{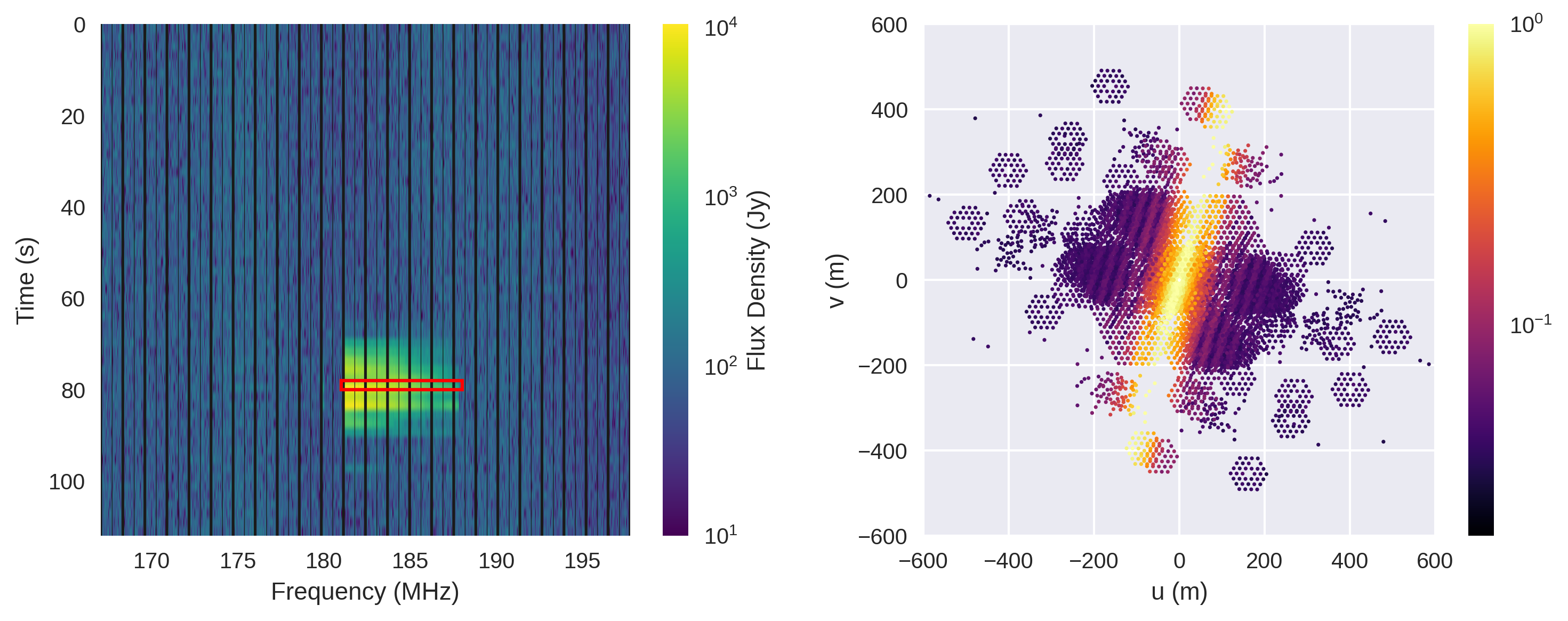}
    \caption{MWA OBSID 1252945816. The selected interval begins 78 s into the observation and covers 181-188 MHz.}
    \label{fig:additional_example1}
\end{figure*}

\begin{figure*}
    \centering
    \includegraphics[width=0.9\linewidth]{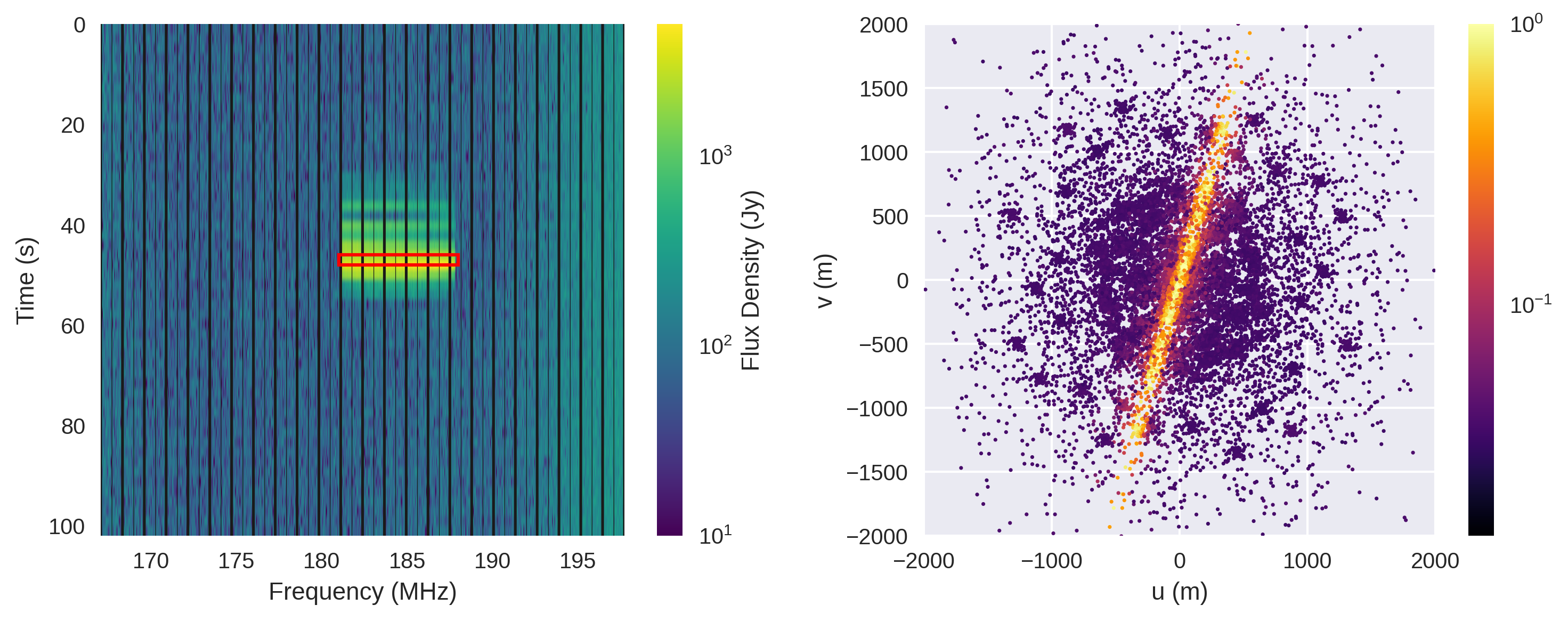}
    \caption{MWA OBSID 1092761680. The selected interval begins 46 s into the observation and covers 181–188 MHz.}
    \label{fig:additional_example2}
\end{figure*}

\begin{figure*}
    \centering
    \includegraphics[width=0.9\linewidth]{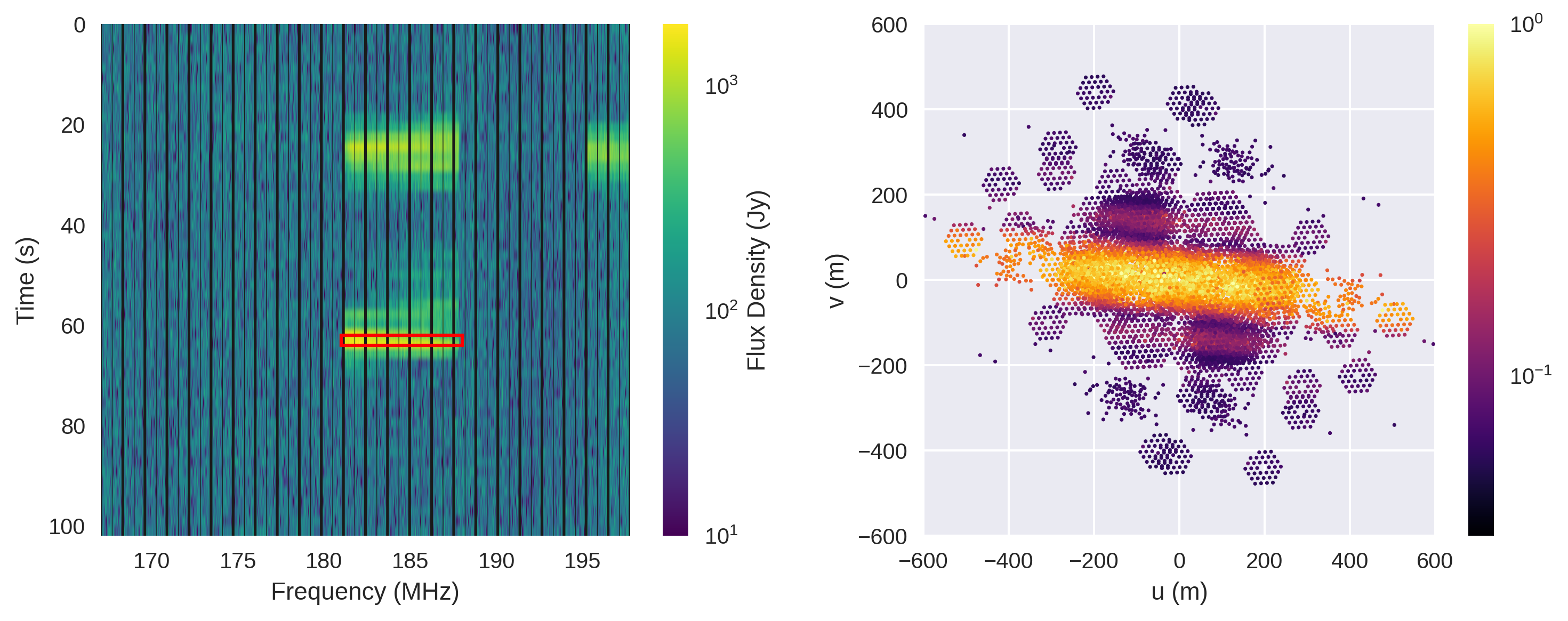}
    \caption{MWA OBSID 1160575504. The selected interval begins 62 s into the observation and covers 181–188 MHz.}
    \label{fig:additional_example3}
\end{figure*}

\begin{figure*}
    \centering
    \includegraphics[width=0.9\linewidth]{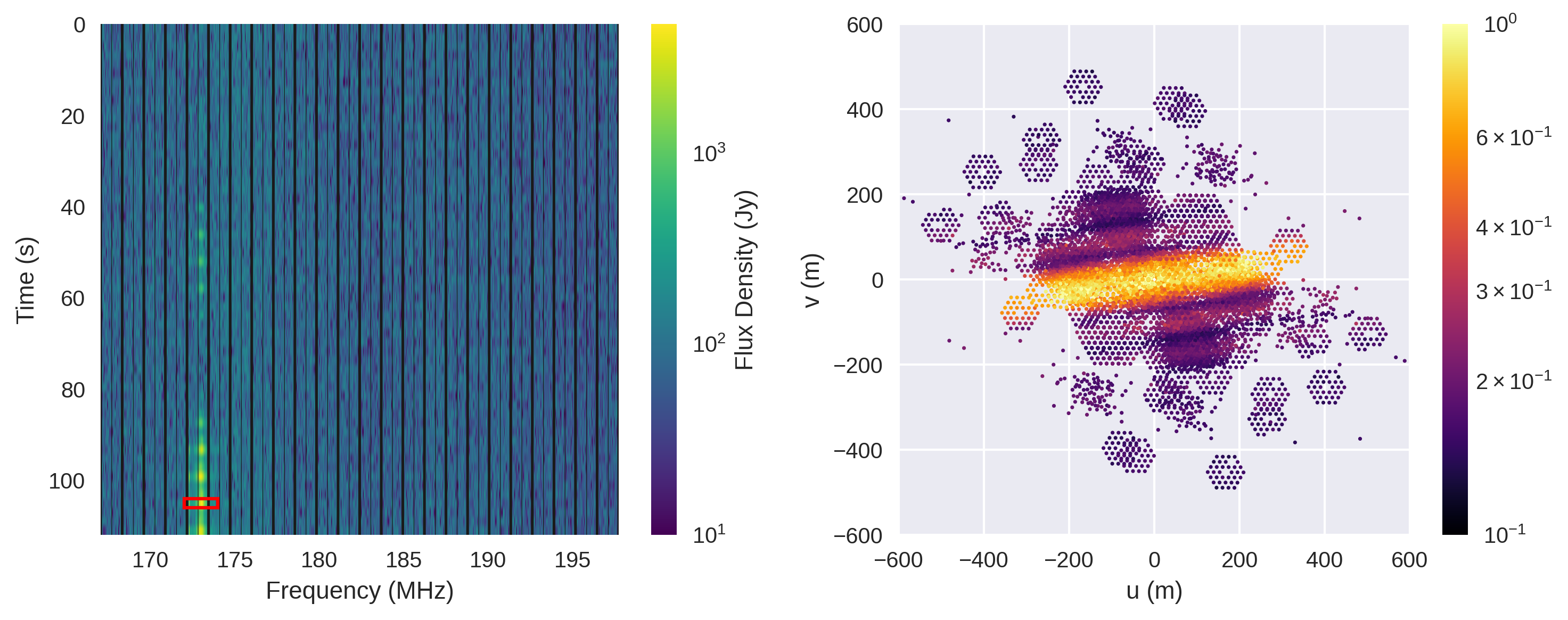}
    \caption{MWA OBSID 1255099560. The selected interval begins 104 s into the observation and covers 172-174 MHz.}
    \label{fig:additional_example4}
\end{figure*}

\begin{figure*}
    \centering
    \includegraphics[width=0.9\linewidth]{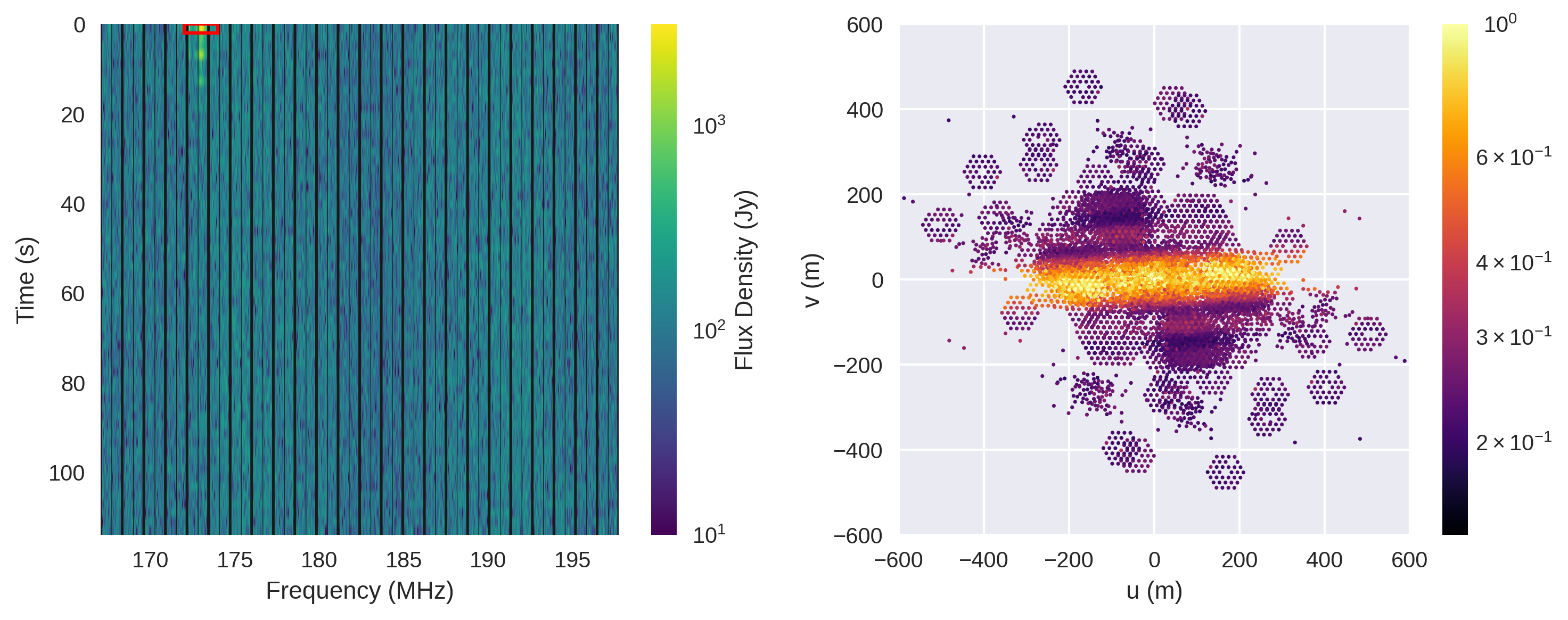}
    \caption{MWA OBSID 1255099680. The selected interval begins 0 s into the observation and covers 172-174 MHz.}
    \label{fig:additional_example5}
\end{figure*}

\end{document}